# Curvature of Energy Dispersions in Condensed Matter Quantum Systems

Tishya Patel[1], Yusen Ye[2] and Tharindu Fernando[*]

[1]*Ahmedabad International School, Shidhnath Mandir Marg, Bodakdev, Ahmedabad, GJ, India 380054*
[2]*Department of Materials Science and Engineering, University of Washington, Seattle, Washington, USA, 98195*
[*]*Department of Physics, University of Washington, Seattle, Washington, USA, 98195*

[*]tharindu@uw.edu

**Abstract:** Graphene, a two-dimensional material with tunable electronic properties, holds significant importance in condensed matter physics and material science. In this study, we analyze the curvature of graphene's ground-state energy dispersion band by examining its Gaussian and mean curvature under varying on-site potential values, M, using graphene's Hamiltonian under periodic boundary conditions. We diagonalized the Hamiltonian matrix to obtain eigenvalues on a discretized grid for the Brillouin zone in momentum space, constructing the ground-state energy dispersion band. Surface differentials are then calculated to subsequently calculate Gaussian and mean curvature at each point in the grid using the definitions of the first and second fundamental forms. These values were collated to form curvature plots, and we plotted the logarithm of maximum Gaussian and mean curvature values from each plot and against the corresponding values of the logarithm of M. After analyzing these plots, we observe that the both the Gaussian and mean curvature remain approximately the same until the on-site potential is equal or greater than the nearest-neighbor hopping parameter of the Graphene Hamiltonian, and then starts beginning to decline. Therefore, this study offers insights into studying the graphene Hamiltonian further and shows that studying the mean and Gaussian curvature of energy dispersions can prove to be a viable method of studying quantum systems. Further directions may be to topologically study other Hamiltonians.

## INTRODUCTION

Condensed matter physics is heavily connected to multiple other fields, such as materials science, engineering, and nanotechnology. Studies in condensed matter physics focus on developing an understanding about how particles (for example, electrons) interact or behave, and how their interplay collectively gives rise to a material's properties. Understanding these interactions using theoretical models can help us predict the behavior of these materials. One concept used in condensed matter physics are the energy dispersion relations, which describe the allowed energy states of particles as they move through the material. This helps us understand how these particles behave and hence improves our understanding about the material's electrical and optical properties.

Therefore, accurately studying these energy dispersions – which can be represented by *energy value surfaces* (more commonly known as *band diagrams*) – is crucial. Finding new methods to study these is also quite essential, as they may make calculations more convenient or accurate. In this paper we study the energy surfaces of graphene using their Gaussian and mean curvature. Graphene is a two-dimensional material, which may potentially replace silicon-based transistors and be used to make more efficient capacitors ("super-capacitors") due to its alterability and its tunable electrical properties. Graphene's potential applications in the field of condensed matter make it a material which demands further research. The version of graphene we use is graphene with on-site potential [1][2]: it has energy dispersions that can be modified by changing its structure physically and chemically [11], which makes it a very versatile material. It also has a special energy dispersion at its Dirac points (where electrons act like massless Dirac particles), and most importantly it has a simple 2 by 2 Hamiltonian matrix (for which it is will be easy to create the energy eigenvalue surface, for which we will find the mean and Gaussian curvature).

In general, energy surfaces geometry hasn't been studied extensively using the Gaussian and mean curvature [6][7][8]. Hence, this perspective may help us further investigate relationships between the curvature of energy dispersions and properties of quantum materials. In this paper, we aim to bridge these concepts from differential geometry to condensed matter physics and show that there are correlations between the geometry of the energy surfaces and physically observable properties.

## BACKGROUND THEORY

The time-independent Schrödinger equation may be used to describe the behavior of electrons:

$$E\Psi = \hat{H}\Psi$$

In the linear eigenvalue equation above, the allowed energies of the system are given by the eigenvalues $E$ of the Hamiltonian operator matrix $\hat{H}$. $\Psi$ is the quantum mechanical wavefunction of the electron (a complex-valued vector).

In periodic systems, one can use Fourier transformation [3][4] to convert the real-space system into momentum space (i.e. k-space). In doing so, one solves the Hamiltonian in the k-space, and the eigenvalues and eigenvectors are labeled by their momentum, k, as per Bloch's theorem [9].

Using the above approach, a Hamiltonian matrix for graphene with on-site potential $M$ [1] is:

$$H(k,M) = -t \sum_{\delta_i} \begin{pmatrix} M & \cos(k.\delta) + i\sin(k.\delta) \\ \cos(k.\delta) - i\sin(k.\delta) & -M \end{pmatrix}$$

where the sum $\Sigma$ is taken over the nearest-neighbor vectors of the honeycomb/hexagonal lattice: $\delta_1 = \frac{a}{2}(1,\sqrt{3})$, $\delta_2 = \frac{a}{2}(1,-\sqrt{3})$, and $\delta_3 = -a(1,0)$, $a$ is the lattice constant, $t$ is the nearest-neighbor hopping parameter [1][2], and $k = (k_x, k_y)$ represents momentum space (i.e. k-space). Note that for real graphene, $M = 0$ (i.e. zero on-site potential). $M$ can be made non-zero by applying external stimuli such as an electric field or stack the graphene on top of hexagonal boron nitride (hBN). See Figure 1 for a visualization of the system. In Figure 1, the vectors $a_i$ represent a possible set of (one orientation) lattice vectors for the hexagonal lattice of graphene (in real space). These vectors represent the shortest possible translations required to replicate the hexagonal primitive cell of graphene's crystalline lattice. When a Fourier transform from real space of these vectors (and graphenes lattice) to k-space is carried out, these vectors are transformed to their respective reciprocal lattice vectors, i.e. $b_i$.

The lattice vectors of graphene are defined as: $\vec{a_1} = a\left(\frac{\sqrt{3}}{2}, \frac{-1}{2}, 0\right)$ and $\vec{a_2} = a\left(\frac{\sqrt{3}}{2}, \frac{1}{2}, 0\right)$ (also, $\vec{a_3} = (0,0,1)$ which isn't a lattice vector but is crucial for calculating the reciprocal lattice vectors). The reciprocal lattice vectors can then be calculated using the following equations:

$$\vec{b_1} = \frac{2\pi(\vec{a_2} \times \vec{a_3})}{\vec{a_1} \cdot (\vec{a_2} \times \vec{a_3})}, \quad \vec{b_2} = \frac{2\pi(\vec{a_3} \times \vec{a_1})}{\vec{a_1} \cdot (\vec{a_2} \times \vec{a_3})}$$

After substituting the vectors and simplifying the expressions above, the reciprocal lattice vectors end up $\vec{b_1} = \frac{4\pi}{\sqrt{3}a}\left(\frac{1}{2}, \frac{-\sqrt{3}}{2}\right)$ and $\vec{b_2} = \frac{4\pi}{\sqrt{3}a}\left(\frac{1}{2}, \frac{\sqrt{3}}{2}\right)$. The Brillouin zone (BZ) is the rhombus shape formed by these vectors. The BZ can be subdivided into two similar equilateral triangles, in the center of these triangles are the K and K´ points. In Graphene these points are also the Dirac points.

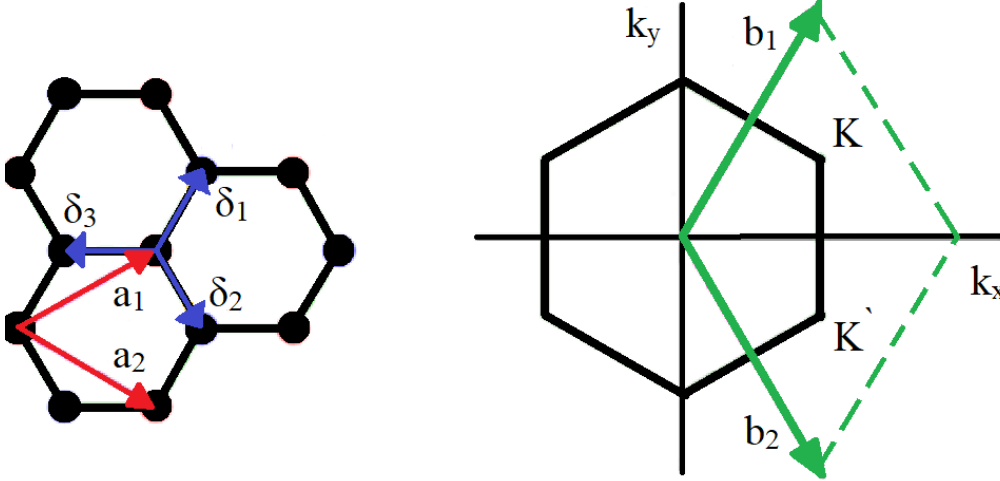

**FIGURE 1.** [13] This shows the nearest-neighbor vectors $\delta_i$, lattice vectors $a_i$, reciprocal lattice vectors $b_i$ and the Dirac points $K, K'$ of the Brillouin zone. The diagram on the right is found by a Fourier transform from real space (diagram on the left) to momentum space.

To find the energies for this matrix, we use the characteristic polynomial to solve the eigenvalue equation. Since the matrix is $2 \times 2$ of rank 2, we get two eigenvalues $E \pm$ at each k-point, which correspond to the ground state energy and excited state energy. For this model, it is known that as the value for $M$ is increased, the band gap between the ground state energy band and excited state energy also increases. We later verify this using our numerical simulations. We do this since it can be very difficult to get analytic solutions/equations (and in our case, is not worth the effort since we want to study complex differential geometric quantities, which will result in even more complicated equations). Additionally, this allows us to develop a framework to easily study other systems, as we simply need to replace the Hamiltonian with an appropriate model for another system.

Since we've found the eigenvalues at each k-point, we get surfaces for each of the two quantum states. In this work, we use differential geometric techniques to study only the ground state eigenvalue surface. This study could be extended to the excited state surface as well, but we limit the scope to the ground state for simplicity, and because, in this model, it is known that the excited state energies are the ground state energies multiplied by $-1$, and so will effectively have similar curvatures (potentially up to a sign difference). For the ground state surface, which we henceforth represent by $s(x, y)$ for $x \equiv k_x$ and $y \equiv k_y$, we calculate the coefficients of the first and second fundamental forms ($E, F, G$, and $L, M, N$ respectively) to find the Gaussian curvature $K$ (a measure of the surface's intrinsic curvature) and mean curvature $H$ (a measure of extrinsic curvature)[6]. These quantities are defined by:

$$K = \frac{LN - M^2}{EG - F^2} \text{ and } H = \frac{LG - 2MF + NE}{2(EF - F^2)},$$

where the definitions of the coefficients of the first fundamental form are:

$$E = \frac{\partial s(x,y)}{\partial x} \cdot \frac{\partial s(x,y)}{\partial x}, \quad F = \frac{\partial s(x,y)}{\partial x} \cdot \frac{\partial s(x,y)}{\partial y}, \text{ and } G = \frac{\partial s(x,y)}{\partial y} \cdot \frac{\partial s(x,y)}{\partial y}$$

Above, the . represents the dot product, and $\frac{\partial}{\partial x}$ represents the partial derivative with respect to variable $x$. Note that $K$ is a label for both the $K$ point and the Gaussian curvature, and $E$ is the label for the energy and a first fundamental form coefficient. We will clarify which definition we will use if not clear from context.
The second fundamental form coefficients are:

$$L = \frac{\partial}{\partial x}\left(\frac{\partial s(x,y)}{\partial x}\right) \cdot n, \quad M = \frac{\partial}{\partial x}\left(\frac{\partial s(x,y)}{\partial y}\right) \cdot n, \text{ and } N = \frac{\partial}{\partial y}\left(\frac{\partial s(x,y)}{\partial y}\right) \cdot n$$

Above, $n$ is the unit normal vector found by taking the cross-product $\times$ of the partial derivates of the surface with respect to $x$ and $y$, and then divided by the cross-product's magnitude | |:

$$n = \frac{\frac{\partial s(x,y)}{\partial x} \times \frac{\partial s(x,y)}{\partial y}}{\left|\frac{\partial s(x,y)}{\partial x} \times \frac{\partial s(x,y)}{\partial y}\right|}$$

## METHODOLOGY

For our simulations, we use Python [5] to calculate the energies $E$. We discretize the k-space Brillouin zone. We create a regular $700 \times 700$ grid of evenly spaced k-points and solve the eigenvalue Schrödinger equation at each point to numerically construct the energy surfaces. For a sanity check, we plot the ground state and excited state energy surfaces $E(k_x, k_y) \equiv s(x, y)$. Figure 2 shows what these surfaces look like for $M = 0$ (naturally occurring in graphene) and non-zero $M = 5$. As mentioned earlier, we see that the eigenvalues are negatives of each other (symmetric across the z-axis). For $M = 0$, we see the bands touch at the Dirac points $K$ and $K'$.

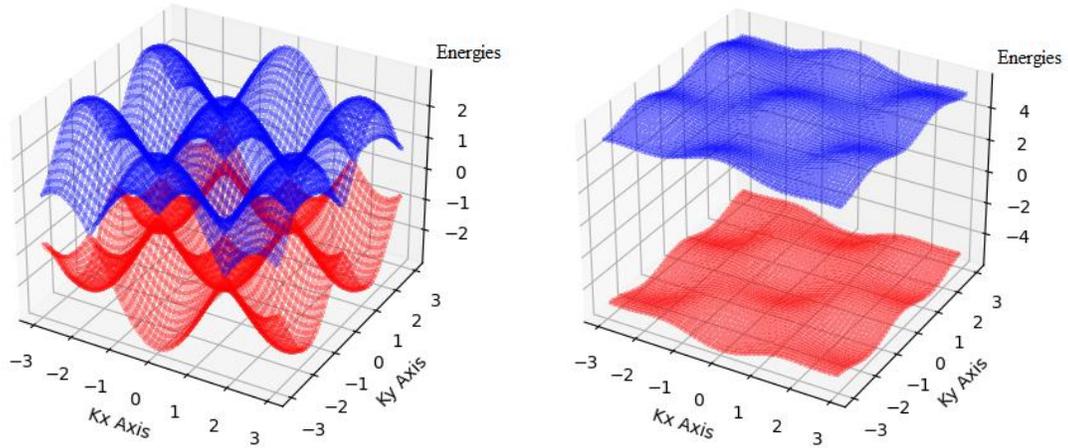

**FIGURE 2.** Values for energy surfaces for $M = 0$ (left) and $M = 5$ (right). The z-axis gives the energies of the ground state (bottom) and excited state (top). The $M = 0$ case shows Dirac points, where the energies are degenerate at $K$ and $K'$ (i.e. surfaces touch). The surfaces do not touch when $M \neq 0$.

We store the ground state energies $E$ in an array $(k_x, k_y, E)$ and use this to calculate the Gaussian and mean curvatures. We calculate the partial derivatives using numerical differentiation. We use the forward differences method between two consecutive k-points (in the x and y directions respectively for the $\partial_x$ and $\partial_y$ derivatives, as implemented using the following equations:

$$\partial_x (s(x,y)) \approx \frac{s(x+h,y)-s(x-h,y)}{2h}$$

$$\partial_y (s(x,y)) \approx \frac{s(x,y+k)-s(x,y-k)}{2k}$$

We calculate the curvatures over the k-space grid for a variety of $M$ values $(-1000, -100, -10, -5, -1, 0, 1, 5, 10, 100, 1000)$ and study correlations between these curvatures and the energy surfaces.

The code we used in this work may be found in our GitHub repository [Patel, T., & Fernando, T. (2023). Curvature-of-graphene-energy-surface (Version 1.0.0) [Computer software]. https://doi.org/10.5281/zenodo.1234].

## RESULTS

After the Gaussian and mean curvature plots for the different M value energy bands are made results can be drawn. In Figure 3 below plots of Gaussian and Mean curvature for M= -1 and M=1 can be seen. Analyzing them, we can see that the K (Gaussian curvatures) plots are exactly equivalent to each other, same with the H (Mean curvature) plots. From this we can infer that the sign of M does not have any impact on the curvature of the energy surface.

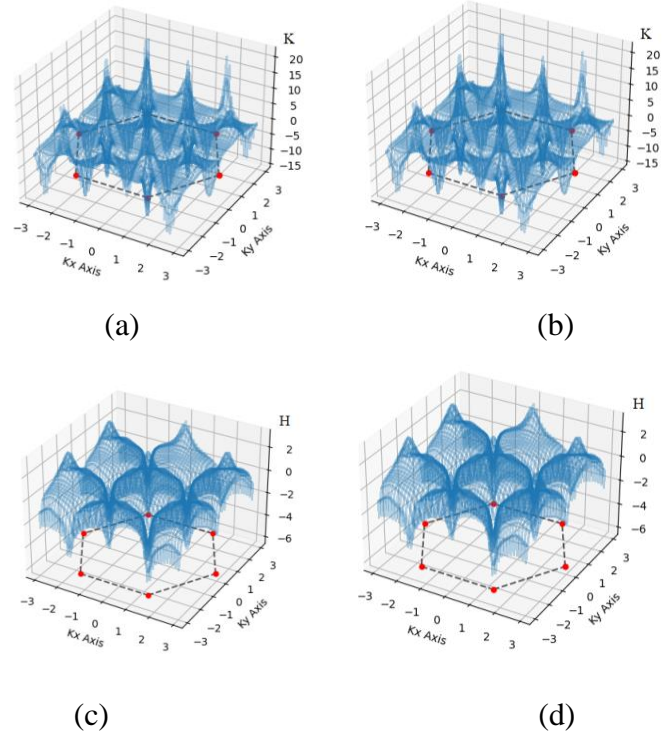

(a)  (b)

(c)  (d)

**FIGURE 3.** These plots represent the Gaussian and Mean curvature for M's value being 1 or -1, (a) show the Gaussian curvature for M=1 while (b) shows Gaussian curvature for M=-1. (c) is the plot for Mean curvature for M=1 and (d) the plot for Mean curvature for M=-1 is positive.

As |M| approaches 0 the magnitude of K and H both increase. The plots below show this correlation as well. M=0 (like in graphene) the curvature diverges when comparing to the plots for M=1) (at the Dirac points) however the Gaussian curvature is much more extreme when compared to the mean curvature, this would make sense since the intrinsic curvature should feel much more drastic than the extrinsic curvature, (Ant on the energy band sees as drastic change at Dirac point, but when we look from outside, it is not such a big curvature). The Gaussian curvature at the Dirac points diverges, this is because the energy dispersion surface forms a conical structure near these points, and the Dirac points represent the tip of the cone where K is undefined. Since an undefined K value cannot be represented by a computer-generated plot, it is approximately displayed by a large value while plotting.

In the K plot of M=0 the curvature seems to only be at the Dirac points and seems to be flat at the rest of the areas, however, restricting this K range show that there is some curvature in that area as shown in the plots below.

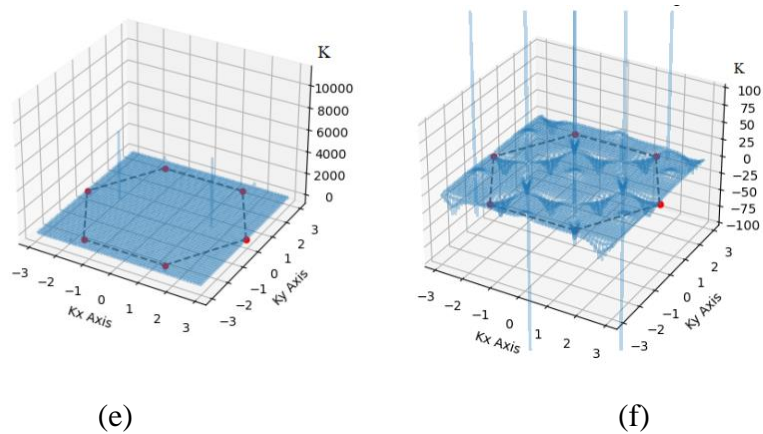

(e)  (f)

**FIGURE 4.** Both (e) and (f) are K plots for M=0 however in (f) the K-range has been limited and hence the curvature in the other areas of the plot is also visible not only at the Dirac points, in (e) the curvature at the Dirac points overwhelms the curvature at the other points so it only seems like there is only curvature at the Dirac points and the rest of the plot is flat.

To further examine the correlation between curvature and energy dispersion the plots below have been generated, which display natural log of maximum curvature against the natural log of its corresponding M value. Natural log has been used to decrease the large range disparity between the two quantities (i.e. curvatures and M).

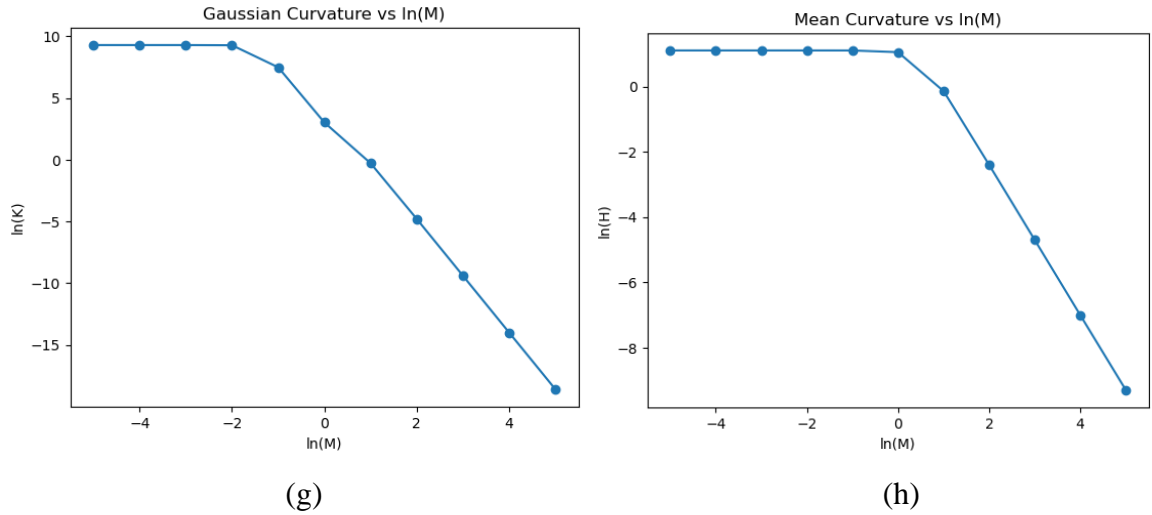

(g)  (h)

**FIGURE 5.** The plot (g) represents the graph of ln K (y-axis) against ln M (x-axis). Plot (h) represents the graph of ln H (y-axis) against ln M (x-axis).

In the plots we see that for (both K and H) when M is significantly smaller than 1 (i.e. $ln(M) \ll 0$) the curvature of the energy dispersion band structure doesn't increase much (negligible), with change in M, forming the flat linear segments in the plots above. At $ln(M) = 0$ $(M = 1)$ the graph begins to decline and continues to do so in a linear manner for any $M > 1$. To mathematically determine strength of the linear relation in these segments, we can determine the Pearson correlation coefficient for the segments individually. The formula below represents the calculation for the correlation coefficient.

$$R = \frac{\sum (x_i - \overline{x})(y_i - \overline{y})}{\sqrt{\sum (x_i - \overline{x})^2 \sum (y_i - \overline{y})^2}}$$

$x_i$ is a particular value of $ln(M)$ from the values of M for which curvature values have been calculated and plotted as band structures, while its corresponding $y_i$ is the max value of $ln(Max\ Curvature\ Value)$ for that band structure (of that value of M). $(\overline{x})$ and $(\overline{y})$ is the Mean value of the values of all $x_i$ and $y_i$ respectively.

We calculate R using Python for: $ln(M)$ against $ln(K)$, and $ln(M)$ against $ln(H)$ (separately calculating R for $ln(M) > 0\ and\ ln(M) < 0$, for the two segments of the plots) both.

The R value for the ln(K) against ln(M), for $ln(M) > 0$ is 0.9974, and ln(M)< 0 is 0.9738. In the ln (Mean curvature) against ln(M) plot, R for $ln(M) > 0$ is 0.9969 and ln(M)< 0 is 0.9989. This displays that the graph segments have very strong linear correlation.

Since the nearest neighbor hopping parameter for the matrix, we have used is t=1, these results are justified. At $ln(M) > 0\ i.e.\ M \gg 1$ the mass term dominates and the bandstructure flattens, therefore displaying a low Gaussian and Mean curvature, which continually reduce as M is increased. For M<<1 (ln(M) < 0) the bandgap between the excited state energy dispersions and ground state energy dispersions starts to become neglible and the hopping parameter dominates, making the graphene act like zero on-site potential graphene and the physics of that quantum system could be explained by the gapless model of graphene. This is also why we see a consistent yet extremely high curvature for these band structures.

At ln(M) = 0 or M=1, the mass term and hopping parameter almost become equal and hence there is competition between them and hence we see that the decline in the curvature values begin from M=1.

# CONCLUSION AND FUTURE DIRECTIONS

After analyzing the results, it can be stated that there are studying the topology of the energy eigenvalue band (i.e. the curvature) could be a viable way of studying material properties. These properties arise from the interactions of subatomic particles within the system, which can be studied through energy dispersion bands.

Analyzing energy dispersion bands using mean and Gaussian curvature, while varying on-site potential, displays that curvature changes that with change in band-gap (which changes the properties of material), specifically it reduces with higher band-gap. This is most evident in M=0 plots, where curvature diverges at Dirac points where electrons act as massless particles, which influences electron mobility and conductivity, whereas in systems with higher band-gap the curvature converges and gets significantly lower as band-gap increases.

Overall, studying the curvature of these energy bands can prove to be a viable new method for examining quantum systems and predicting their properties.

Future directions could involve applying other Hamiltonians to extend this topological and curvature-based analysis to different material systems, potentially enriching our understanding of material properties in condensed matter physics. As we further explore this method, topological study may become a more established and effective approach within materials science and condensed matter physics, providing a bridge between theoretical models and observable properties.

## ACKNOWLEDGMENTS

We thank Mittul Patel, Ritu Patel for helping conduct this research possible.

## AUTHOR CONTRIBUTIONS

T.P. served as the primary author, implementing the project under the guidance provided by T.F and Y.Y.. T.F. conceived the project, provided foundational guidance, and supported initial implementation. Y.Y. assumed supervisory responsibilities in the latter stages, overseeing the completion and finalization of the work

## REFERENCES

1. *Franz Utermohlen, Tight-Binding Model for Graphene September 12, 2018.*
2. *Castro Neto, A, et al. The Electronic Properties of Graphene.*
3. *Kittel, Charles. Introduction to Solid State Physics. John Wiley & Sons, 1971.*
4. *Harris, C. R., Millman, K. J., van der Walt, S. J., Gommers, R., Virtanen, P., Cournapeau, D., ... Oliphant, T. E. (2020). Array programming with NumPy. Nature, 585, 357–362. https://doi.org/10.1038/s41586-020-2649-2*
5. *Jia, Yan-Bin. Gaussian and Mean Curvatures *. 2020.*
6. *B. O'Neill. Elementary Differential Geometry. Academic Press, Inc., 1966.*
7. *A. Pressley. Elementary Differential Geometry. Springer-Verlag London, 2001*
8. *Watanabe, H. A Proof of the Bloch Theorem for Lattice Models. J Stat Phys 177, 717–726 (2019). https://doi.org/10.1007/s10955-019-02386-1*
9. *Patel, T., & Fernando, T. (2023). Curvature-of-graphene-energy-surface (Version 1.0.0) [Computer software]. https://doi.org/10.5281/zenodo.1234*
10. *Schober, Patrick MD, PhD, MMedStat; Boer, Christa PhD, MSc; Schwarte, Lothar A. MD, PhD, MBA. Correlation Coefficients: Appropriate Use and Interpretation. Anesthesia & Analgesia 126(5):p 1763-1768, May 2018. | DOI: 10.1213/ANE.0000000000002864*
11. *Geim, A. K. "Graphene: Status and Prospects." Science, vol. 324, no. 5934, 18 June 2009, pp. 1530–1534, https://doi.org/10.1126/science.1158877. Accessed 12 Dec. 2019.*
12. *Olver, Peter J. Introduction to Partial Differential Equations. Cham, Switzerland, Springer, 2020.*